\documentclass[12pt]{article}

 \newread\testifexists
 \def\GetIfExists #1 {\immediate\openin\testifexists=#1
     \ifeof\testifexists\immediate\closein\testifexists\else
     \immediate\closein\testifexists\input #1\fi}

 \usepackage{gthstyle}\usepackage{amsfonts}
 \usepackage{amssymb}
 \immediate\openin\testifexists=e
     \ifeof\testifexists\immediate\closein\testifexists\else
     \immediate\closein\testifexists\input e\fipsf

 \def\Bbb#1{\setbox0=\hbox{$\tt #1$}  \copy0\kern-\wd0\kern .1em\copy0}

 \def\bbf#1{\setbox0=\hbox{$#1$} \kern-.025em\copy0\kern-\wd0
         \kern.05em\copy0\kern-\wd0 \kern-.025em\raise.0433em\box0}

 \immediate\openin\testifexists=a
     \ifeof\testifexists\immediate\closein\testifexists\else
     \immediate\closein\testifexists\input a\fimssym.def  

 \newcommand{\crlb}[1]{\label{#1}\\[2pt]}
 \newcommand{\eela}[1]{\quad\hbox{\scriptsize{#1}}\label{#1}\end{eqnarray}}
 \newcommand{\eelb}[1]{\label{#1}\end{eqnarray}}
 
 \newcommand{\newsecb}[2]{\section{#1}\label{#2}\setcounter{equation}{0}}
 
 \newcommand{\nolabels} {\def\eel{\eelb} \def\crl{\crlb} \def\newsecl{\newsecb}}

\newcommand\publishversion{\nolabels\setlength{\textheight}{9in}\setlength{\oddsidemargin}{0in}
    \setlength{\textwidth}{6.3in}\setlength{\topmargin}{-0.1in}}

 \def\a{\alpha}      \def\b{\beta}   \def\g{\gamma}      
 \def\d{\delta}      \def\D{\Delta}  \def\e{\varepsilon} 
 \def\k{\kappa}      \def\l{\lambda} \def\L{\Lambda}     \def\m{\mu}
 \def\f{\phi}            \def\vv{\varphi}    \def\n{\nu}
 \def\j{\psi}                 \def\s{\sigma}

 \def\w{\omega}        

    \def\LL{{\mathcal L}} \def\OO{{\mathcal O}} \def\DD{{\mathcal D}}\def\NN{{\mathcal N}}
 \def\pa{\partial} \def\ra{\rightarrow}
  
 \def\dd{{\rm d}}     

 \def\qu{\ {\buildrel {\displaystyle ?} \over =}\ }
 
 \def\iss{\ =\ }

 \def\fract#1#2{{\textstyle{#1\over#2}}}
 \def\ffract#1#2{\raise .2 em\hbox{$\scriptstyle#1$}\kern-.3em/
                 \kern-.2em\lower .15 em \hbox{$\scriptstyle#2$}}
 
 \def\half{\fract12} \def\quart{\fract14} 

 \def\ex#1{e^{\textstyle#1}}

 \newcommand{\tl}[1]{\tilde{#1}}              \newcommand{\Tr}{{\mbox{Tr}}\,}
                     \newcommand{\fn}{\footnote}
 \newcommand{\nn}{\nonumber\\[2pt]}             \newcommand{\nm}{\nonumber}
 \newcommand{\be}{\begin{eqnarray}}             \newcommand{\ee}{\end{eqnarray}}
 \newcommand{\bi}[1]{\begin{itemize}\item[#1]}         
       \newcommand{\ei}{\end{itemize}}
 \newcommand{\eqn}[1]{(\ref{#1})}

\publishversion
\begin{document} \begin{titlepage}

\title{
\vskip 20mm \Large\bf The Conformal Constraint in \\ Canonical Quantum Gravity}

\author{Gerard 't~Hooft}
\date{\normalsize Institute for Theoretical Physics \\
Utrecht University \\ and
\medskip \\ Spinoza Institute \\ Postbox 80.195 \\ 3508 TD Utrecht, the Netherlands \smallskip \\
e-mail: \tt g.thooft@uu.nl \\ internet: \tt http://www.phys.uu.nl/\~{}thooft/}

\maketitle

\begin{quotation} \noindent {\large\bf Abstract } \medskip

{
{Perturbative canonical quantum gravity is considered, when coupled to a renormalizable model for matter fields. It is
proposed that the functional integral over the dilaton field should be disentangled from the other integrations over
the metric fields. This should generate a conformally invariant theory as an intermediate result, where the conformal
anomalies must be constrained to cancel out. When the residual metric is treated as a background, and if this
background is taken to be flat, this leads to a novel constraint: in combination with the dilaton contributions, the
matter lagrangian should have a vanishing beta function. The zeros of this beta function are isolated points in the
landscape of quantum field theories, and so we arrive at a denumerable, or perhaps even finite, set of quantum theories
for matter, where not only the coupling constants, but also the masses and the cosmological constant are all fixed, and
computable, in terms of the Planck units.}}

\end{quotation}

  \vfill \flushleft{October 30, 2010}

\end{titlepage}

\eject
\newsecl{Introduction}{intro}  \def\mat{\mathrm{\,mat}}\def\EH{\mathrm{\,EH}}\def\kin{\mathrm{kin}}
\def\mass{\mathrm{mass}} \def\intt{\mathrm{int}}
In a previous paper\cite{GtHconfgrav}, it was argued that the functional integral in canonical quantum gravity,
 \be \int\DD g_{\m\n}\DD\f^\mat\,\ex{i(S^\EH(g_{\m\n})+ S^\mat(g_{\m\n},\f^\mat))}\ , \eel{canonical}
where \(S^\EH\) is the Einstein-Hilbert action and \(S^\mat\) is the action of the matter fields,
here abbreviated as \(\f^\mat\), should be considered to be taken in two steps:
 \be g_{\m\n}\equiv\w^2\hat g_{\m\n}\ ;\quad \det(\hat g_{\m\n})=-1\ ,\quad \int\DD g_{\m\n}=\int\DD\hat g_{\m\n}\int\DD\w\ ,
 \eel{metricsplit}
and the integral over the dilaton field \(\w\) should be taken together with the integrations over the matter fields
\(\f^\mat\).

Rewriting the Einstein-Hilbert action (including a possible cosmological constant) in terms of \(\w\) and \(\hat
g_{\m\n}\), one finds, in four dimensions,
 \be S^\EH=\int\dd^4x{1\over 2\k^2}\Big(\hat R\,\w^2+6\,\hat
 g^{\m\n}\pa_\m\w\pa_\n\w-2\L\w^4\Big)\ , \eel{EHsplit}
where \(\k^2=8\pi G_N\), and \(\hat R\) is the Ricci scalar associated to \(\hat g_{\m\n}\). It is convenient to split
the lagrangian of the matter fields into conformally invariant kinetic parts, mass terms, and interaction terms. In a
simplified notation (later we will be more precise), one has
 \be \f^\mat&=&\{A_\m(x), \j(x),\bar\j(x),\vv(x)\}\ ;\crl{matterfields}
 \LL^\kin&=&-\quart\hat g^{\m\a}\hat g^{\n\b}F_{\m\n}F_{\a\b}-\half\hat
 g^{\m\n}\pa_\m\vv\,\pa_\n\vv-\fract1{12}\hat R\vv^2-\bar\j \g^\m\hat D_\m\j\ ;\crl{conformalL}
 \LL^\mass&=&-\half m_s^2\w^2\vv^2-\bar\j\,\w\m_d\,\j\ . \eel{massL}
Here, \(F_{\m\n}=\pa_\m A_\n-\pa_\n A_\m\); the kinetic term for the Dirac fields is shorthand for the corresponding
expression using a vierbein field \(\hat e_\m^a\) for the metric \(\hat g_{\m\n}\) with its associated connection
field. \(m_s\) stands short for the scalar masses and \(m_d\) for the Dirac masses. The term \(\fract1{12}\hat R\vv^2\)
could be removed by a field redefinition, but is kept here for convenience, making the scalar lagrangian conformally
invariant.

Interaction between the matter fields must now be written in the form
 \be\LL^\intt=-\fract 1{4!}\l\vv^4-\bar\j y_i\vv_i\j-\fract1{3!} g_3\vv^3\w\ , \eel{interL}
where the Yukawa couplings \(y_i\) could be matrices in the indices labeling the fermion species, and \(\l\) and
\(g_3\) could be 4- and 3-index tensors in the scalar field indices.

If now we rescale the \(\w\) field:
 \be \w(x)={\k\over\sqrt 6}\,\tl\w(x)\ , \eel{omegarescale}
we notice two things. First, the action for the \(\tl\w\) field is now nearly identical to the kinetic term for the
\(\vv\) field, both having the same conformal dimension:
 \be\LL^\EH=\half\hat g^{\m\n}\pa_\m\tl\w\,\pa_\n\tl\w+\fract1{12}\hat R\tl\w^2-\fract1{36}\k^2\L\w^4\ , \eel{omegatildeL}
and secondly, the mass terms turn into \emph{conformally invariant quartic coupling terms} between matter fields and
the \(\tl\w\) field:
 \be\LL^\mass=-\half\,{\tl\k^2 m_s^2}\,\tl\w^2\vv^2 -{\tl\k m_d}\,\bar\j\,\tl\w\,\j\ ,\eel{confmass}
where \(\tl\k= \k/\sqrt6 =\sqrt{\fract43\pi G_N}\) has the dimension of an inverse mass. Also the scalar 3-field
coupling, which originally was not conformally invariant, now turns into a conformally invariant 4 field coupling (see
Eq.~\eqn{interL}):
 \be -\fract 1{3!}\tl\k g_3 \vv^3\tl\w\ , \eel{threefield}
and a new quartic interaction term for the \(\tl\w\) field is generated by the cosmological constant:
 \be -\fract16\tl\k^2\L\,\tl\w^4\ . \eel{cosmoint}

It is important to observe that the lagrangian \eqn{omegatildeL} for the dilaton field \(\tl\w\) has an overall sign
opposite to that of ordinary scalar fields \(\vv\). For any other field theory, this would be disastrous because it
would violate causality. Here however, the unconventional sign is a necessary consequence of the canonical structure of
the theory. Since it is an overall sign, it has no net effect on the Feynman rules; this we will exploit by rotating
the field in the complex plane:
 \be\tl\w(x)\equiv i\eta(x)\ , \eel{omegarotate}
so that the new field \(\eta(x)\) will be indistinguishable from other scalar fields, with one important exception: the
conformal interaction terms from the original mass terms, Eq.~\eqn{confmass}, as well as the interaction from the
original 3-field interaction, Eq.~\eqn{threefield}, get unconventional factors -1 or \(i\).

Another important thing to observe is that there is no term at all in the lagrangian that could serve as a kinetic term
for the \(\hat g_{\m\n}\) field\fn{Of course, this term reappears if one substitutes \(\eta=-i+\OO(\k\tl\eta)\), where
\(\tl\eta\) is a small oscillating field, by expanding \(\hat R\).}. Any such kinetic terms should arise solely from
higher order effects due to the interactions with the matter fields. Naturally, since the entire lagrangian now is
conformally invariant, we should expect the effective lagrangian for \(\hat g_{\m\n}\) to be conformally invariant as
well. It is emphasized in Ref.~\cite{GtHconfgrav} however that, as is well known \cite{Duff}, this conformal invariance
is mutilated by conformal anomalies.

In this paper, we now propose to postpone any attempts to describe the functional integrals over \(\hat g_{\m\n}\). Not
only do we have anomalies there, but also there are difficulties with unitarity and Landau ghosts.\fn{In
Ref.~\cite{Mannheim},\cite{BM} claims are made that unitarity can be restored. This however requires the integration
contours to be rotated in the complex plane such that \(\hat g_{\m\n}\) becomes complex. Such approaches are
interesting and may well serve as good starting points for generating promising theories, but they will not be pursued
here.} As was explained in Ref.\cite{GtHconfgrav}, the effective interactions with matter and dilaton fields generate
an action for \(\hat g_{\m\n}\) that largely coincides with the familiar conformally invariant action obtained from the
square of the Weyl curvature, \emph{but with an infinite numerical coefficient}, which would have to be renormalized.
The difficulties associated to that are sufficient reason for us now to postpone this sector of the theory entirely.

At first sight, one may well find logical objections to such a procedure: why not also first integrate over \(\hat
g_{\m\n}\) before examining whether the amplitudes obtained obey conformal constraints? We argue however that the
\(\hat g_{\m\n}\) integration is very different from the rest; \(\hat g_{\m\n}\) determines the location of the local
light cones, so that it determines the causal relationships between points in space-time. It may well be that quantum
interference of states with light cones at different places will require treatments that differ in essential ways from
the standard functional integral.

In any case, it is worth investigating what happens if we follow this procedure. The implications are quite remarkable,
as we will show.

\newsecl{Renormalization}{renorm}
Let us assume that the matter fields \(\f^\mat\) consist of Yang-Mills fields \(A_\m^a\), Dirac fields
\(\bar\j,\ \j\) and scalar fields \(\vv\), the latter three sets being in some (reducible or irreducible,
chiral or non chiral) representation of the local Yang-Mills gauge group. For brevity, we will write complex
scalar fields as pairs of real fields, and if Weyl or Majorana fermions occur, the Dirac fields can be
replaced by pairs of these.\fn{A single Weyl or Majorana fermion then counts as half a Dirac field.} Let us
rewrite the lagrangian for matter interacting with gravity more precisely than in the previous, introductory
section:
 \be \LL(\hat g_{\m\n},\eta,\f^\mat)&=& -\quart G^a_{\m\n}G^a_{\m\n}-\bar\j\,\hat\g^\m \hat D_\m\,\j-\half \hat g^{\m\n}
 (D_\m\vv\,D_\n\vv+\pa_\m\eta\,\pa_\n\eta) \nn && -\fract1{12}\hat R(\vv^2+\eta^2)-V_4(\vv)-iV_3(\vv)\eta+\half\tl\k^2
 m_i^2\eta^2\vv_i^2-\tl\L\eta^4 \nn &&-\bar\j(y_i\vv_i+iy_i^5\g^5\vv_i+i\tl\k m_d\eta)\j\ , \eel{fullL}
 where \(G_{\m\n}\) is the (non Abelian) Yang-Mills curvature, and \(D_\m\) and \(\hat D_\m\) are covariant derivatives containing the Yang-Mills fields; \(\hat\g_\m\) and \(\hat D_\m\)
also contain the vierbein fields and connection fields associated to \(\hat g_{\m\n}\); the Yukawa couplings
\(y_i\), \(y_i^5\) and fermion mass terms \(m_d\) are matrices in terms of the fermion indices. The scalar
self interactions, \(V_3(\vv)\) and \(V_4(\vv)\) must be a third and fourth degree polynomials in the fields
\(\vv_i\):
 \be V_4(\vv)&=&\fract1{4!}\l\vv^4\iss\fract1{4!}\l^{ijk\ell}\vv_i\vv_j\vv_k\vv_\ell\ ;\crl{fourfield}
 V_3(\vv)&=&\fract1{3!}\tl g_3^{ijk}\vv_i\vv_j\vv_k\ ,\qquad \tl g_3=\tl\k g_3\ . \eel{threefield1}
In Eq.~\eqn{fullL}, like \(m_d\), also \(m_i^2\d_{ij}\) are mass matrices, in general. Furthermore, \(\tl\L\) stands
for \(\fract16\tl\k^2\L\). Of course, all terms in \eqn{fullL} must be fully invariant under the Yang-Mills gauge
rotations. They must also be free of Adler Bell Jackiw anomalies.\cite{ABJ}

Now that the dilaton field \(\eta\) has been included, the entire lagrangian has been made conformally invariant. It is
so by construction, and no violations of conformal invariance should be expected. This invites us to consider the beta
functions of the theory. Can we conclude at this point that the beta functions should all vanish?

Let us not be too hasty. In the standard canonical theory, matter fields and their ineractions are renormalized. Let us
consider dimensional renormalization, and the associated anomalous behavior under scaling. In \(4-\e\) dimensions,
where \(\e\) is infinitesimal, the scalar field dimensions are those of a mass raised to the power \(1-\e/2\), so that
the couplings \(\l\) have dimension \(\e\). This means that in most of the terms in the lagrangian~\eqn{fullL} the
integral powers of \(\eta\) will receive extra factors of the form \(\eta^{\pm\e}\) or \(\eta^{\pm\e/2}\), which will
then restore exact conformal invariance at all values for \(\e\). If we follow standard procedures, we accept that
\(\eta\) is close to \(-i\), so that singularities at \(\eta\ra0\) or \(\eta\ra\infty\) are not considered to be of any
significance. Indeed, the limit \(\eta\ra0\) may be seen to be the small-distance limit. This is the limit where
gravity goes wrong anyway, so why bother?

However, now one could consider an extra condition on the theory. Let us assume that the causal structure, that is, the
location of the light cones, is determined by \(\hat g_{\m\n}\), and that there exist dynamical laws for \(\hat
g_{\m\n}\). This was seen to be a very useful starting point for a better understanding of black hole
complementarity\cite{GtHcomplement}. The laws determining the scale \(\w(x)\) should be considered to be dynamical
laws, and the canonical theory of gravity itself would support this: formally the functional integral over the \(\eta\)
fields is exactly the same as that for other scalar fields.

In view of the above, we do think it is worthwhile to pursue the idea that the \(\eta\) field must be handled
just as any other scalar component of the matter fields; but then, after renormalization, fractional powers,
in the \(\e\ra0\) limit would lead to \(\log(\eta)\) terms, and these must clearly be excluded.
Renormalization must be done in such a way that no traces of logarithms are left behind. Certainly then, a
scale transformation, which should be identical to a transformation where the fields \(\eta\) are scaled,
should not be associated with anomalies. Implicitly, this also means that the region \(\eta\ra 0\) is now
assumed to be regular. This is the small distance region, so that, indeed, our theory says something
non-trivial about small distances. This is why our theory leads to new predictions that eventually should be
testable. Predictions follow from the demand that all beta functions of the conformal ``theory"~\eqn{fullL}
must vanish.

Note that one set of terms is absent in Eq.~\eqn{fullL}: the terms linear in \(\vv\) and hence cubic in \(\eta\). This,
 of course, follows from the fact that, usually, no terms linear in the scalar fields are needed in the standard matter
 lagrangians; such terms can easily be removed by translations of the fields: \(\vv_i\ra\vv_i+a_i\) for some constants
 \(a_i\). Thus, the classical lagrangian is stationary when the fields vanish: \(\vv=0\) is a classical solution. In
our present notation, this observation is equivalent to the observation that fields may be freely transformed into one
another without modifying the physics. One such transformation is a rotation of one of the scalar fields, say
\(\vv_1\), with the \(\eta\) fields:
  \be \Big\{\matrix{\vv_1 &\ra& \vv_1\,\cosh\a_1+i\eta\,\sinh\a_1 \ ,\cr \eta&\ra&\eta\,\cosh\a_1-i \vv_1\,\sinh\a_1\ ,}
  \eel{etarotate}
where \(\a_1\) stands for the original shift of the field \(\vv_1\). The transformation is taken to be a hyperbolic
rotation because the ``kinetic term" \(-\half(\pa\eta^2+\pa\vv^2)= \half(\pa\tl\w^2-\pa\vv^2)\) in Eq.~\eqn{fullL} has
to be invariant.

In most cases, these transformations need not be considered since terms linear in \(\vv\) will in general not be gauge
invariant.

Notice also that the Yang-Mills fields are not directly coupled to the \(\eta\) field. In the ``classical limit",
\(\eta\ra -i\), we can see why this is so. Since not \(\eta\), but \(\tl\w\) is real, the invariant quantity is
\(\vv^2-\w^2\). Rotating \(\vv\) fields with \(\eta\) fields would therefore be a non-compact transformation, and Yang
Mills theories with non-compact Lie groups usually do not work. There is food for thought here, but as yet we will not
pursue that.

When constructing a complete theory, the next step should be to consider just any configuration of \(\hat g_{\m\n}(\vec
x,t)\), formulate the renormalized theory in this background, and finally consider functional integrals over \(\hat
g_{\m\n}\). Unfortunately, this is still too difficult. In a non-trivial \(\hat g_{\m\n}\) background, there will be
anomalies depending on the derivatives of \(\hat g_{\m\n}\); there are divergences\cite{GtHconfgrav} proportional to
the Weyl curvature squared, which can be seen to correspond to the field combination \( \hat R_{\m\n}^{\ 2}-\fract13
\hat R^2\), and subtracting those leads to new conformal anomalies\cite{Duff}. It was suggested in
Ref.~\cite{GtHconfgrav} to keep the conformal infinity, which would turn gravitons into ``classical" particles, or more
precisely, particles that cannot interfere quantum mechanically, but whether this can be held up as a theory remains to
be seen. To avoid further complications, we now decided to look at the case when \(\hat g_{\m\n}(\vec
x,t)=\eta_{\m\n}\), or, space-time is basically flat, though we just keep the field \(\eta(\vec x,t)\).

\newsecl{The \(\b\) functions}{beta}

Thus, we return to a theory to which all known quantum field theory procedures can be applied, the only new thing being
the presence of an extra, gauge neutral, spinless field \(\eta\), and the perfect local scale invariance of the theory.

We arrived at the lagrangian~\eqn{fullL}, and we wish to impose on it the condition that all its beta functions vanish,
since conformal invariance has to be kept. As the theory is renormalizable, the number of beta functions is always
exactly equal to the number of freely adjustable parameters. In other words: we have exactly as many equations as there
are freely adjustable unknown variables, so that \emph{all coupling constants, all mass terms and also the cosmological
constant,} should be completely fixed by the equations \(\b_i=0\). They are at the stationary points. Masses come in
the combination \(\tl\k\,m_i\) and the cosmological constant in the combination \(\tl\k^2\L\), so all dimensionful
parameters of the theory will be fixed in terms of Planck units.

In principle, there is no reason to expect any of these fixed points to be very close yet not on any of the axes, so
neither masses nor the cosmological constant can be expected to be unnaturally small, at this stage of the theory. In
other words, as yet no resolution of the hierarchy problem is in sight: why are many of the physical mass terms 40
orders of magnitude smaller than the Planck mass, and the cosmological constant more than 120 orders of magnitude? We
have no answer to that in this paper, but we will show that the equations are quite complex, and exotic solutions
cannot be excluded.

The existence of infinitely many solutions cannot be excluded. This is because one can still adjust the
composition and the rank(s) of the Yang-Mills gauge group, as well as one's choice of the scalar and (chiral)
spinor representations.\fn{which of course must be free of Adler Bell Jackiw
anomalies\cite{ABJ}\cite{bardeen}.} These form infinite, discrete sets. However, many choices turn out not to
have any non trivial, physically acceptable fixed point at all: the interaction potential terms \(V(\vv)\)
must be real and properly bounded, for instance. Searches for fixed points then automatically lead to
vanishing values of some or more of the coupling parameters, which would mean that the symmetries and
representations have not been chosen correctly.

Every advantage has its disadvantage. Since all parameters of the theory will be fixed, we cannot apply perturbation
theory. However, we can make judicious choices of the scalar and spinor representations in such a way that the
existence of a fixed point for the gauge coupling to these fields can be made virtually certain. The \(\b\) function
for \(SU(N)\) gauge theories with \(N_f\) fermions and \(N_s\) complex scalars in the elementary representation is
 \be 16\pi^2\,\b(g)&=&-a\,g^3-b\,g^5+\OO(g^7)\ ,\crl{betathree}
 a&=&\fract {11}3 N-\fract23 N_f-\fract16 N_s\ , \crl{betag}
 b&=&\OO(N^2,\ N\,N_f,\ N\,N_s)\ . \eel{betatwoloop}
Choosing one scalar extra, or one missing, we can have \(a\) as small as \(a=\pm\fract16\), while a quick inspection in
the literature\cite{twoloops}\cite{threeloops} reveals that, in that case, \(b\) may still have either sign:
 \be b=\pm\OO(N^2)\ . \eel{twoloopsign}
depending on further details, such as the ratio of fermions and scalars, the presence of other representations, and the
values of the Yukawa couplings. Choosing the sign of \(a\) opposite to that of \(b\), one then expects that a fixed
point can be found at
 \be g^2=-b/a=\OO(1/N^2)\ . \eel{pertfixedpoint}

This, we presume, is close enough to zero that the the following procedure may be assumed to be reliable. Let there be
\(\n\) physical constants, the \(\n^\mathrm{th}\) one being the gauge coupling \(g\), which is determined by the above
equation \eqn{pertfixedpoint}. If we take all other coupling parameters to be of order \(g\) or \(g^2\), then the beta
function equations are reliably given by the one-loop expressions only, which we will give below. Now these are
\(\n-1\) equations for the \(\n-1\) remaining coupling parameters, and they are now inhomogeneous equations, since the
one coupling, \(g^2\), is already fixed. All we have to do now, is find physically acceptable solutions. We already saw
that non-Abelian Yang-Mills fields are mandatory; we will quickly discover that, besides the \(\eta\) fields, both
fermions and other scalar matter fields are indispensable to find any non-trivial solutions.

One trivial, yet interesting solution must be mentioned: \(\NN=4\) super Yang-Mills. We take its lagrangian, and add to
that the \(\eta\) field while postulating that this \(\eta\) field does not couple to the \(\NN=4\) matter fields at
all. Then indeed all \(\b\) functions vanish\cite{susyym}. However, since the \(\eta\) field is not allowed to couple,
the physical masses are all strictly zero, which disqualifies the theory physically. Note, however, that also the
cosmological constant is rigorously zero. Perhaps the procedure described above can be applied by modifying slightly
the representations in this theory, so that a solution with masses close to zero, and in particular a cosmological
constant close to, but not exactly zero, emerges.

The one loop \(\b\) functions are generated by an algebra\cite{GtHalgebra}, in which one simply has to plug the Casimir
operators of the Yang Mills Lie group, the types of the representations, the quartic scalar couplings and the fermionic
couplings. If we take the scalar fields \(\vv_i\) and \(\eta\) together as \(\s_i\), the generic lagrangian can be
written as
 \be\LL=-\quart G_{\m\n}^a G_{\m\n}^a -\half(D_\m\s_i)^2-V(\s)-\bar\j\big(\g D-(S_i+i\g^5P_i)\s_i\big)\j\ , \eel{genL}
where \(\s_i\) and \(\bar\j,\ \j\) are in general in reducible representations of the gauge group, \(D_\m\)
is the gauge covariant derivative, \(V(\s)\) is a gauge-invariant quartic scalar potential, and \(S_i\) and
\(P_i\) are matrices in terms of the fermion flavor indices. Everything must be gauge invariant and the
theory must be anomaly free\cite{ABJ}\cite{bardeen}.

The covariant derivatives contain the hermitean representation matrices \(T^a_{ij}\), \(U^{L\,a}_{\a\b}\) and
\(U^{R\,a}_{\a\b}\):
 \be D_\m\s_i&\equiv&\pa_\m\s_i+iT^a_{ij}A_\m^a\s_j\ ;\crl{scalarcovder}
  D_\m\j_\a&\equiv&\pa_\m\j_\a+i(U^{L\,a}_{\a\b}P^L+U^{R\,a}_{\a\b}P^R)A_\m^a\j_\b\ ;\quad P^{L,R}
  \equiv\fract12(1\pm\g^5)\ . \eel{spinorcovder}
The gauge coupling constant(s) \(g\) are assumed to be included in these matrices \(T\) and \(U\). The operators
\(P^L\) and \(P^R\) are projection operators for the left- and right handed chiral fermions.

The group structure constants \(f^{abc}\) are also assumed to include a factor \(g\), and they are defined by
 \be [T^a,\,T^b]=if^{abc}T^c\ ;\quad [U^{L\,a},\,U^{L\,b}]=if^{abc}U^{L\,c}
 \ ;\quad [U^{R\,a},\,U^{R\,b}]=if^{abc}U^{R\,c}\ .{\quad} \eel{structurecnsts}
Casimir operators \(C_g,\ C_s\) and \(C_f\) will be defined as
 \be f^{apq}f^{bpq}=C^{ab}_g\ ,\quad \Tr(T^aT^b)=C_s^{ab}\ ,\quad \Tr(U^{L\,a}U^{L\,b}+U^{R\,a}U^{R\,b})=C_f^{ab}\ .
 \eel{}

All beta functions are given by writing down how the entire lagrangian~\eqn{genL} runs as a function of the scale
\(\m\)\cite{GtHalgebra}:
 \be{\m\pa\over\pa\m}\LL&=&\b(\LL)\iss{1\over 8\pi^2}\D\LL\ , \crl{betaLdef} 
\D\LL&=& -\quart G_{\m\n}^aG_{\m\n}^b(\fract{11}3C_g^{ab}-\fract16C_s^{ab}-\fract23C_f^{ab})-\D V-\bar\j(\D S_i+i\g^5\D
P_i)\s_i\,\j\ . \eel{deltaL} Here,
 \be\D V&=&\quart V_{ij}^2-\fract32 V_i(T^2\s)_i+\fract34(\s T^aT^b\s)^2\ +\nn
 &&\s_iV_j\Tr(S_iS_j+P_iP_j)-\Tr(S^2+P^2)^2+\Tr[S,P]^2\ , \eel{deltaV}
 where \(\ V_i=\pa V(\s)/\pa\s^i\ ,\quad V_{ij}\iss\pa^2V(\s)/\pa\s_i\pa\s_j \) .

\noindent It is convenient to define the complex matrices \(W_i\) as
 \be W_i=S_i+iP_i\ , \eel{Wmatrices}
Then, \be \D W_i&=&\quart W_kW_k^*W_i+\quart W_iW_k^*K_k+W_kW_i^*W_k\ -\nn
 &&-\ \fract32 (U^R)^2W_i-\fract32 W_i(U^L)^2+W_k\Tr(S_kS_i+P_kP_i)\ . \eel{deltaW}

If now we write the collection of scalars as \(\{\s_i= \vv_i,\ \s_0=\eta\}\), taking due notice of the factors \(i\) in
all terms odd in \(\eta\), we can apply this algebra to compute all \(\b\) functions of the lagrangian~\eqn{fullL}.

The values of the various \(\b\) functions depend strongly on the choice of the gauge group, the representations, the
scalar potential function and the algebra for the Yukawa terms, and there are very many possible choices to make.
However, the signs of most terms are fixed by the algebra~\eqn{deltaL}---\eqn{deltaW}. By observing these signs, we can
determine which are the most essential algebraic constraints they impose on possible solutions. As we will see, they
are severely restrictive.

\newsecl{Adding the dilaton field to the algebra for the \(\b\) functions.}{scalaralgebra}

Consider the dilaton field \(\eta\) added to the lagrangian, as in Eq.~\eqn{fullL}. This requires extending the indices
\(i,\,j,\,\dots\) in the lagrangian \eqn{genL} to include a value \(0\) referring to the \(\eta\) field. The unusual
thing is now that the terms odd in \(\eta\) are purely imaginary, while all terms in Eq.~\eqn{fullL} are of dimension
4. Let us split off this special component. In Eq.~\eqn{genL}, we then write

 \be V(\s)&=&\tl\L\eta^4-\half m_i^2\eta^2\vv_i^2+iV_3(\vv)\eta+V_4(\vv)\ ; \crl{fullV}
 S_0=im_d&;&\quad P_0=0\ ;\qquad S_i=y_i\ ;\qquad  P_i=y^5_i\ . \eel{fullW}
Here, \(m_d\) is the fermionic mass matrix, \(y_i\) are matrices representing the scalar Yukawa couplings, and
\(y_i^5\) are the pseudoscalar Yukawa coupling matrices. \(m_i^2\d_{ij}\) is the scalar mass matrix, which is allowed
to have negative eigenvalues, so we allow the Higgs mechanism to take place. We henceforth choose modified Planck units
by setting \(\tl\k^2=\fract43\pi G_N=1\).

Note, that in the Standard Model, there is no gauge-invariant Dirac mass matrix and no gauge-invariant cubic
scalar interaction, so there \(m_d\) and \(V_3\) are zero, but we will need to be more general.

We write \(W=S+iP\), and now \(\tl W=S-iP\). The algebra \eqn{deltaL} --- \eqn{deltaW} is then found to become
 \be &\D V(\s)\ =\ \D V^0(\vv) &(a)\nn
 &-\half V_{3i}^2-V_3\vv_i\Tr(m_d y_i)+\quart(\,m_i^2\vv_i^2\,)^2 &(b)\nn
 &+i\eta\,\bigg(\ -\fract32 V_{3i}(T^2\vv)_i+\half V_{3ij}V_{4ij}+\vv_iV_{3j}\Tr(y_iy_j+y_i^5y_j^5)+V_{4i}\Tr(m_dy_i)
  &\crl{extalgebra}
  &-4\Tr(m_d(y_i\vv_i)^3)+2\Tr([m_d,\,y^5_i\vv_i][y_j\vv_j,\,y^5_k\vv_k])-2\Tr((y_i^5\vv_i)^2(m_dy_j+y_jm_d)\vv_j)&\nn
  &-2V_{3i}m_i^2\vv_i -V_3\Tr m_d^2-m_j^2\vv_j^2\vv_i\Tr(m_dy_i)\ \bigg)&(c)\nn
 &+\eta^2\,\bigg(\ -\quart V_{3ij}^2-V_{3i}\Tr(m_dy_i)-\half m_i^2V_{4ii}^2+2\Tr((m_d\,y_i\vv_i)^2)+
 4\Tr(m_d^2(y_i\vv_i)^2)&\nn
 & +2\Tr(m_d^2(y_i^5\vv_i)^2-\Tr([m_d,\,y^5_i\vv_i]^2)-m_j^2\vv_i\vv_j\Tr(y_iy_j+y^5_iy^5_j)+\fract32
 m_i^2\vv_i(T^2\vv)_i&\nn
 &+2m_i^4\vv_i^2+\Tr(m_d^2)m_i^2\vv_i^2-6\tl\L m_i^2\vv_i^2\ \bigg)&(d)\nn
 &+i\eta^3\,\bigg(-\half m_i^2V_{3ii}-m_i^2\vv_i\Tr(m_dy_i)+4\Tr(m_d^3\,y_i\vv_i)+4\tl\L\vv_i\Tr(m_dy_i)\ \bigg)&(e)\nn
 &+\eta^4\,\bigg(\,36\tl\L^2-4\tl\L\,\Tr(m_d^2)+\quart\sum_im_i^4-\Tr(m_d^4)\,\bigg)\ ,&(f)\nn
 &&\nm\ee
where \(V_{3i}=\pa V_3/\pa \vv_i\), etc, and we apply summation convention: double indices are summed over starting
with 1, except the index in the scalar mass matrix \(m^2_i\), which is only summed over if it occurs twice elsewhere as
well, or if this is explicitly indicated. \(\D V^0\) is the expression that we already had, in Eq.~\eqn{deltaV}.

The beta coefficients for the Yukawa couplings follow from adding the index 0 to Eq.~\eqn{deltaW}:
 \be \D W_i&=&\D W_i^0-\quart(m_d^2\,W_i+W_i\,m_d^2)+m_d\tl W_i\,m_d-m_d\Tr(m_d\,S_i)\ ,\qquad{ }\crl{deltayuk}
 -i\D W_0\iss\D m_d&=&-\fract32 m_d^3-m_d\Tr(m_d^2)+\quart(y_k^2+{y_k^5}^2)m_d+\quart m_d(y_k^2+{y_k^5}^2)\nn
 &&+\ y_k\,m_d\,y_k-y_k^5\,m_d\,y_k^5+y_k\Tr(y_k\,m_d)+i\g^5y_k^5\,\Tr(y_k\,m_d)\ , \eel{deltadiracmass}
where \(\D W_i^0\) stands for the standard \(\b\) function for the corresponding dimension 4 interaction terms.

Before demanding that the \(\b\) functions all vanish, we observe that we can still allow for an infinitesimal field
transformation of the form \eqn{etarotate} in the original lagrangian. This adds to the counter terms:
 \be &\d V(\s)\iss i\a_i\bigg(\eta\displaystyle{\pa V(\s)\over\pa\vv_i}-\vv_i{\pa V(\s)\over\pa\eta}\bigg)& \nn
&=\qquad\a_iV_3\vv_i\ +\ i\eta\,\a_i\bigg(V_{4i}+m_j^2\vv_j^2\vv_i\,\bigg)&\nn
 & +\ \eta^2\,\a_i\,(-V_{3i})\ +\ i\eta^3\,\a_i\bigg(-m_i^2\vv_i-4\tl\L\vv_i\,\bigg)\ , \eel{fieldtrf}
and a similar rotation in the Yukawa couplings. This can be used to eliminate the term (\ref{extalgebra}~e) by
adjusting \(\a_i\); it corresponds to a field shift in the non-gravitational case. In most cases, however, such as in
the Standard Model, the terms in (\ref{extalgebra}~e) are forced to vanish anyhow due to gauge invariance. In a similar
way, an infinitesimal chiral rotation among the fermions can be used to eliminate the last term in
Eq.~\eqn{deltadiracmass}.

Thus, after term (e) has been made to vanish by hand, the demand that all \(\b\) functions vanish, for all values of
\(\eta\), applies in particular to the terms in Eq.~(\ref{extalgebra}~a --- d) and (f), and to Eqs.~\eqn{deltayuk} and
\eqn{deltadiracmass}.

We have already assumed that the non-Abelian Yang Mills field coupling(s) \(g\) have a small but non-vanishing fixed
point. Through the effects of the group matrices \(T^a,\ U^{La}\) and \(U^{Ra}\), the coupling(s) \(g\) determine the
values of the other parameters, by as many coupled non-linear equations as there are unknowns. It follows that, in this
theory, we \emph{must} have non-Abelian Yang-Mills fields. In contrast, Abelian \(U(1)\) components are not allowed
since those do not have fixed points close to the origin.

Next, let us consider the requirement that the term (f) in Eq.~\eqn{extalgebra} vanishes:
 \be 36\tl\L^2=\Tr(m_d^4)+4\tl\L\,\Tr(m_d^2)- \quart\sum_im_i^4 \ . \eel{cosmocancel}
The r.h.s. of this equation resembles a supertrace. Since its sign must be positive, we read off right away that there
\emph{must} be fermions. If furthermore we like to have a very small or vanishing cosmological constant \(\L\), we
clearly need that the sum of the fourth power of the Dirac fields (approximately) equals the sum of the fourth powers
of the masses of the real scalar particles divided by 4.

Can we do without the scalar fields \(\vv_i\)? Eq.~\eqn{cosmocancel} would have a solution, although the cosmological
constant would come out fairly large. However, now there is only one more equation to consider:
Eq.~\eqn{deltadiracmass}, with all Yukawa couplings \(y_i\) and \(y_i^5\) replaced by 0. That gives:
 \be \fract32 m_d^3+m_d\Tr(m_d^2)\qu 0\ . \eel{noscalars}
Whenever \(m_d\) has a real, non vanishing eigenvalue, this would imply that the trace of \(m_d^2\) is negative, an
impossible demand. Therefore, our theory also \emph{must} have scalars \(\vv_i\), besides the dilaton field \(\eta\).

It appears that in today's particle models not only the cosmological constant \(\tl\L\) but also the mass terms are
quite small, in the units chosen, which are our modified Planck units. Also, if there is a triple scalar coupling,
\(V_3(\vv)\), it appears to be small as well. This is the hierarchy problem, for which we cannot offer any solution
other than suggesting that we may have to choose a very complex group structure --- as in the landscape scenarios often
proposed in superstring theories. Perhaps, the small numbers in our present theory are all related.

If the masses are indeed all small, then the only large terms in our equations are the ones that say how the coupling
constants and masses run with scale. Our theory suggests that they might stop running at some scale; in any case, a
light Higgs particle indeed follows from the demand that the Higgs self coupling is near an UV fixed point.

The author did not (yet) succeed in finding a physically interesting prototype model with a non-trivial fixed point;
this is a very complex, but interesting technical problem. Let us briefly set out a strategy.

We search for a solution where all mass terms, and of course also the cosmological constant, are small. Start with a
theory that has nearly, but not quite, a set of \(\b\) functions that vanish at one-loop. Assume that it has a fixed
point near the origin. It is known that Eqs.~\eqn{deltaL}---\eqn{deltaW} allow this. For simplicity, let us assume that
there are no triple scalar couplings, \(V_3=0\), and no gauge invariant scalars, so that the terms (\ref{extalgebra}~c)
and (e) are forbidden by gauge invariance. If we deviate slightly from the fixed point, term (b) dictates that we must
be at a point where the \(\b\) function for the scalar self couplings is negative. This gives us a first guess for
\(m_i^4\), the absolute values of the scalar masses, but not the signs of \(m_i^2\).

Knowing the approximate values of the Yukawa couplings \(y_k\) and \(y^5_k\) allows us to fix the Dirac mass matrix
\(m_d\) using Eq.~\eqn{deltadiracmass}. Since we want these masses to be small, we must assume that the pseudoscalar
Yukawa couplings largely cancel against the scalar ones in this equation. Then, Eq.~\eqn{deltayuk} can be obeyed by
moving slightly away from the original fixed point in that direction.

The only remaining equation is then the vanishing of term (d), the running of the scalar mass-squared terms. Knowing
\(V_{4ii}^2\) and \(\Tr(m_d^2)\), and assuming that \(m_i^4\) is very small, then gives us an equation for \(m_i^2\).
Notice that its sign was still free, so that there is some freedom here. Various further attempts to refine this
procedure may well lead to interesting models with fixed points. We do note that, apparently, relatively large
pseudoscalar Yukawa couplings \(y_i^5\) are wanted. Also, we are talking about primary, gauge invariant Dirac masses,
which have to be there due to the term (\ref{extalgebra}~f), while they do not occur in the presently known Standard
Model.

\newsecl{Discussion.}{disc}

Our theory derives constraints from the fact that matter fields interact with gravity. The basic assumption could be
called a new version of relativity: the scalar matter fields should not be fundamentally different from the dilaton
field \(\eta(\vec x,t)\). Since there are no singular interactions when a scalar field tends to zero, there is no
reason to expect any singularity when \(\eta(\vec x,t)\) tends to zero at some point in space-time. Standard gravity
theory does have singularities there: this domain refers to the short distance behavior of gravity, which is usually
considered to be ``not understood". What if the short distance behavior of gravity and matter fields is determined by
simply demanding the absence of a singularity? Matter and dilaton then join smoothly together in a perfectly
conformally invariant theory. This, however, only works if all \(\b\) functions of this theory vanish: its coupling
parameters must be at a fixed point. There are only discrete sets of such fixed points. Many theories have no fixed
point at all in the domain where physical constants are real and positive --- that is, stable. Searching for non
trivial fixed points will be an interesting and important exercise.

Indeed, \emph{all} physical parameters, including the cosmological constant, will be fixed and calculable in terms of
the Planck units. This may be a blessing and a curse at the same time. It is a blessing because this removes all
dimensionless freely adjustable real numbers from our theory; everything is calculable, using techniques known today;
there is a strictly discrete set of models, where the only freedom we have is the choice of gauge groups and
representations. It is difficult to tell how many solutions there are; the number is probably infinite.

This result is also a curse, because the values these numbers have in the real world is strange mix indeed: the range
of the absolute values cover some 122 orders of magnitude:
 \be \tl\L=\OO(10^{-122})\ ;\qquad \m^2_\mathrm{Higgs}\approx 3.10^{-36}\ . \eel{numvalues}
The question where these various hierarchies of very large, or small, numbers come from is a great mystery called the
``hierarchy problem". In our theory these hierarchies will be difficult to explain, but we do emphasize that the
equations are highly complex, and possibly theories with large gauge groups and representations have the potential to
generate such numbers.

Our theory is a ``top-down" theory, meaning that it explains masses and couplings at or near the Planck domain. It will
be difficult to formulate any firm predictions about physics at energies as low as the TeV domain. Perhaps we should
expect large regions on a logarithmic scale with an apparently unnatural scaling behavior. There is in principle no
supersymmetry, although the mathematics of supersymmetry will be very helpful for constructing the first non-tivial
models.

What is missing furthermore is an acceptable description of the dynamics of the remaining parts \(\hat g_{\m\n}\) of
the metric field. In Ref.~\cite{GtHconfgrav}, it was suggested that this dynamics may be non quantum mechanical,
although this does raise the question how \(\hat g_{\m\n}\) can back react on the presence of quantum matter. Standard
quantum mechanics possibly does not apply to \(\hat g_{\m\n}\) because the notion of energy is absent in a conformal
theory, and consequently the use of a hamiltonian may become problematic. A hamiltonian can only be defined after
coordinates and conformal factor have been chosen, while this is something one might prefer not to do. The author
believes that quantum mechanics itself will have to be carefully reformulated before we can really address this
problem.

Our theory indeed is complex. We found that the presence of non-Abelian Yang-Mills fields, scalar fields and spinor
fields is required, while \(U(1)\) gauge fields are forbidden (at least at weak coupling, since the \(\b\) function for
the charges here is known to be positive). Because of this, one ``prediction" stands out: there will be magnetic
monopoles, although presumably their masses will be of the order of the Planck mass.

Finally, there is one other firm prediction: the constants of nature will indeed be truly constant. Attempts to
experimentally observe variations in constants such as the finestructure constant or the proton electron mass ratio,
with time, or position in distant galaxies, are predicted to yield negative results.

\section*{Acknowledgements}
The author benefited from discussions with C.~Kounnas and C.~Bachas at the ENS in Paris.

\end{document}